\documentclass[aps,showkeys,showpacs,twocolumn]{revtex4}
\usepackage{mathrsfs}
\usepackage{amssymb}
\usepackage{amsmath}
\usepackage{bm}
\usepackage{graphics}
\usepackage{natbib}

\begin{document}
\title{Topological properties in the non-Hermitian tetramerized Su-Schrieffer-Heeger lattices}

\author{Jia-Rui Li$^1$}
%\author{Shu-Feng Zhang$^2$}
\author{Lian-Lian Zhang$^1$}%\email{zhanglianlian@mail.neu.edu.cn}
\author{Wei-Bin Cui$^2$}
\author{Wei-Jiang Gong$^1$}\email{gwj@mail.neu.edu.cn}

\affiliation{1. College of Sciences, Northeastern University, Shenyang 110819, China\\
2. State Key Laboratory of Rolling and Automation, Northeastern University, Shenyang 110819, China}
\date{\today}

\begin{abstract}
In this paper, we study the topological properties of the non-Hermitian Su-Schrieffer-Heeger (SSH) lattice by periodically introducing onsite imaginary potentials in the manner of ($i\gamma_1$, $-i\gamma_2$, $-i\gamma_1$, $i\gamma_2$) where $\gamma_1$ and $\gamma_2$ are the imaginary potential strengths. Results show that by changing the lattice to a tetramerized non-Hermitian system, such imaginary potentials induce the nontrivial transition of the topological properties of the SSH system. First, the topologically-nontrivial region is extended, followed by the non-Hermitian spontaneous breaking of the anti-$\cal PT$ symmetry. In addition, new edge state appears, but its locality is different from the state induced by the Hermitian SSH lattice. If such potentials are strong enough, the bulk states of this system can become purely imaginary states. We believe that these imaginary potentials play special roles in modulating the topological properties of the non-Hermitian SSH lattice.
\end{abstract}
\maketitle

\bigskip
\section{introduction}
During the past years, topological phases have attracted extensive attention in the fields of quantum mechanics and condensed matter physics\cite{1,2,3}, since they represent an exotic form of matter with geometrical origins and emerge without any symmetry breaking. Moreover, topological phases are characterized by global properties of quantum systems rather than local orders, which makes them immune to local perturbations\cite{4}. This benefits emergent topological excitations as the basic states for quantum computation. After the first discovery of topological phases, comprehensive theoretical and experimental studies have been carried to create, classify, and comprehend these exotic phases. Until now, various types of systems have been reported to display topological properties, including photonic\cite{5,6}, solid\cite{7,8}, acoustic\cite{9,10}, atomic\cite{11,12,13} and electronic\cite{14,15} systems.
\par
Among the great variety of topological systems, the well-known Su-Schrieffer-Heeger (SSH) model\cite{16}, a one-dimensional (1D) lattice with alternating hopping coefficients, is one of the most basic and important models for describing the band topology\cite{17,18,19}. Its bulk-boundary correspondence\cite{20} shows that under periodic boundary condition, the winding number, as a topological invariant, distinguishes two topologically distinct regimes determined by the ratio of two hopping coefficients. Under the open boundary condition, localized edge states are found at both ends of the SSH chain. Importantly,  the winding numbers determine the number of edge states. Due to these properties, the SSH models have attracted extensive attention\cite{21,22,23,24,25,26} and extended SSH models have been investigated from different aspects. For instance, the driven SSH model has been studied by adding periodic modulation to the tunneling or the on-site energy\cite{10,27,28}. The long range hopping SSH model has been investigated\cite{29,30} by introducing long range tunnelings to display its new topological properties.
\par
On the other hand, following the progress of both $\cal PT$-symmetric non-Hermitian quantum physics and optics, the topic of $\cal PT$-symmetric topological quantum physics has been one important research direction\cite{toppt1,toppt2,toppt3,toppt3a}. One reason is because of its underlying new physics, and the other is for that theoretical anticipations can be verified in a relatively short period\cite{YoungsunChoi,Ti01,Ti02,Ti03,before1}. Accordingly, topologically protected $\cal PT$-symmetric interface states have been demonstrated in coupled resonators\cite{toppt4,toppt5}, and $\cal PT$-symmetric non-Hermitian Aubry-Andr\'{e} systems\cite{toppt6} and Kitaev models\cite{toppt7,toppt7a} have been theoretically investigated. Relevant conclusions show that universal non-Hermiticity can alter topological regions\cite{toppt8,toppt9,toppt9a}, but topological properties are robust against local non-Hermiticity. Also, $\cal PT$-symmetric discrete-time quantum walk has been realized, with which edge states between regions with different topological numbers and their robustness to perturbations and static disorder have been observed\cite{LXiao}. Latest report began to focus on the two-dimensional $\cal PT$-symmetric system, which shows that the non-Hermitian two-dimensional topological phase transition coincides with the emergence of mid-gap edge states by means of photonic waveguide lattices with judiciously designed refractive index landscape and alternating loss\cite{Kremer}.
\par
As the simplest topological model, the SSH chain certainly attracts extensive attentions, by considering the $\cal PT$ symmetry different aspects. Zhu $et$ $al$. have studied the $\cal PT$-symmetric non-Hermitian SSH model with two conjugated imaginary potentials at two end sites. They find that the non-Hermitian terms can lead to different effects on the properties of the eigenvalue spectra in topologically
trivial and nontrivial phases. And in the topologically trivial phase, the system undergoes an abrupt transition from the
unbroken $\cal PT$-symmetry region to the spontaneously broken $\cal PT$-symmetry region at a certain potential-magnitude $\gamma_c$\cite{Zhu1}. Following this work, many groups dedicate themselves to investigating the properties of $\cal PT$-symmetric non-Hermitian SSH model, including the spontaneous $\cal PT$-symmetry breaking, topological invariants, topological phase, harmonic oscillation at the exceptional point, and topological end states\cite{SSH1,SSH2,SSH3,SSH4,SSH5,SSH6}. In experiment, the $\cal PT$-symmetry breaking can be observed directly. Moreover, the effects of defects on the topological states in the non-Hermitian SSH system have been already considered, and complicated systems have begun to receive attentions, e.g., the
SSH and Kitaev models and non-Hermitian trimerized lattices\cite{Kitaev,Triple}.

\par
In view of the above researches about the $\cal PT$-symmetric non-Hermitian SSH system, we would like to consider the anti-$\cal PT$-symmetric structure can also be realized by alternating the application manner of gain and loss, and in this system new physics mechanisms will be induced. In fact, anti-$\cal PT$-symmetric systems have attracted attentions in recent years\cite{Jin}. In the present work, we propose a more complicate system, i.e., the non-Hermitian tetramerized SSH lattice in which the onsite imaginary potentials are introduced in the manner of ($i\gamma_1$, $-i\gamma_2$, $-i\gamma_1$, $i\gamma_2$) with $\gamma_{1(2)}$ being the strength of the imaginary potentials. Our purpose is to clarify the topological phase transition. Note that due to the abilities of ultracold-atom or optical-lattice systems to implement the non-Hermitian SSH models, our considered system is easy to realize experimentally. As a consequence, calculation results show that such imaginary potentials cause the SSH lattice to show the anti-$\cal PT$ symmetry, and they induce the nontrivial transition of the topological properties. On the one hand, the topologically-nontrivial region is extended, followed by the non-Hermitian spontaneous symmetry breaking. In this process, the new edge state emerges with its alternate locality. Moreover, if the imaginary potentials are strong enough, the bulk states of this system can be thoroughly imaginary. All these phenomena imply interesting properties of the non-Hermitian tetramerized SSH lattice.
\par
The specific content of this paper is organized as follows: In Sec.II, we introduce the Hamiltonian models and calculate the eigenvalues in the momentum space. The symmetries under periodic boundary conditions are also discussed. In Sec. III, the properties of energy spectra under the open boundary condition are explored, as well as the local density of states. Finally, Sec. IV provides a summary and concluding remarks.

\section{Theoretical model}
\begin{figure}
\begin{center}\scalebox{0.11}{\includegraphics{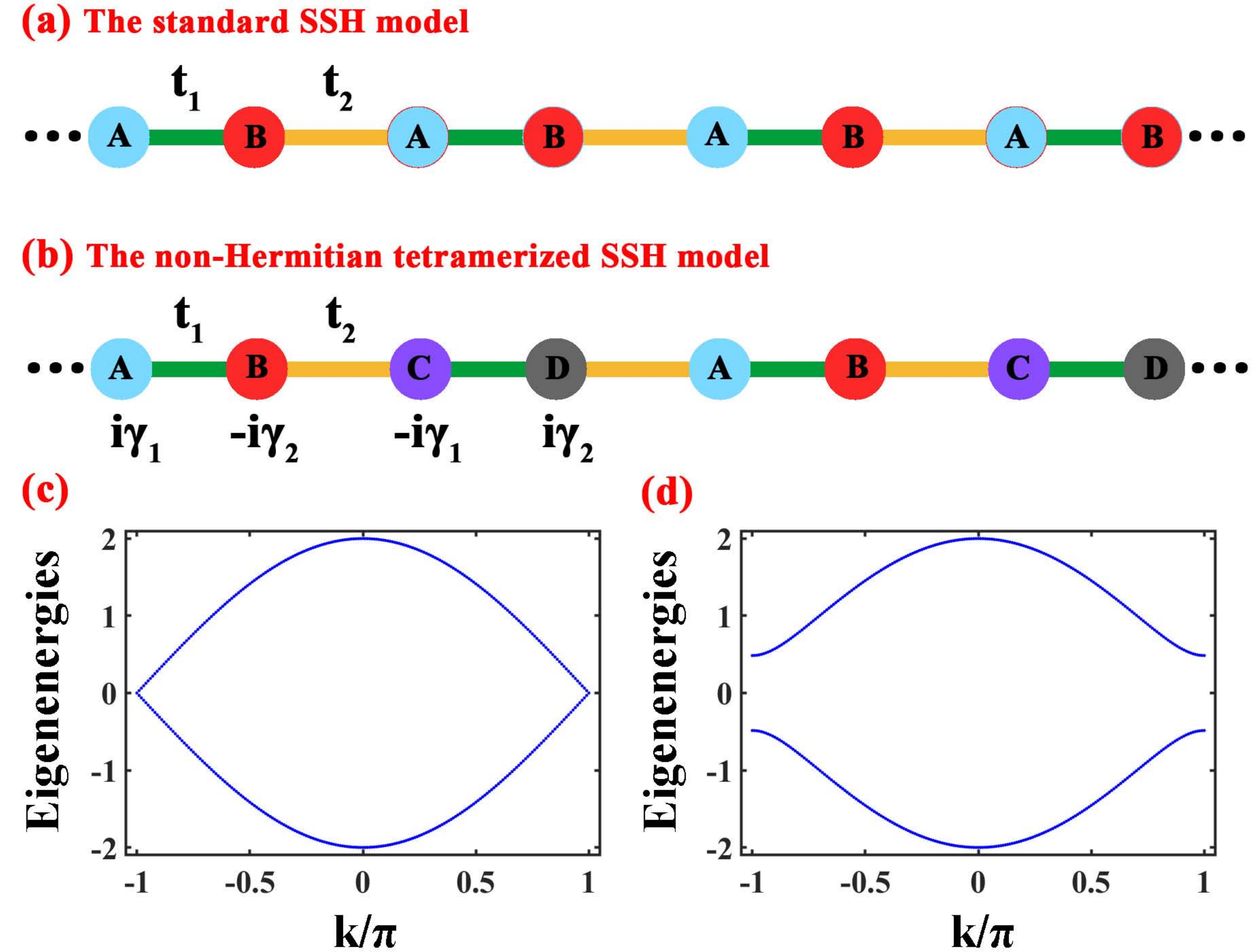}}
\caption{(a) Schematic of a finite section of the Hermitian one-dimensional SSH lattice. A and B represent two different lattices. The green and orange lines represent intracell hopping amplitudes and intercell hopping amplitudes, respectively. (b) Illustration of the non-Hermitian one-dimensional SSH lattice. The imaginary onsite potentials cause the system to become the four-site unit cell lattice $A, B, C$ and $D$. $+i\gamma_1$ and $+i\gamma_2$ represent two different gain potentials. $-i\gamma_1$ and $-i\gamma_2$ describe two different loss potentials. (c)-(d) Band structure of Hermitian tetramerized Su-Schrieffer-Heeger lattices ($\gamma_1=\gamma_2=0$) in  momentum space with (c) $\delta=0.3$, $\theta=0.5\pi$ and (d) $\delta=0.3$, $\theta=0.8\pi$.}
\end{center}
\end{figure}
The non-Hermitian one-dimensional SSH lattice that we consider is illustrated in Fig.1, which describes the spinless particle hopping on the one-dimensional chain composed of $N_{c}$ unit cells.

The Hamiltonian of the non-Hermitian SSH model reads
$H=H_{0}+U$.
$H_{0}$ denotes the Hamiltonian of the one-dimensional Hermitian SSH system [see Fig.1(a)], and it is written as
\begin{eqnarray}
H_{0}&=&\sum_{n=1}^{N_{c}}t_1{a^\dag_{n} b_{n}}+\sum_{n=1}^{N_{c}-1}t_2{b^\dag_{n} a_{n+1}}+h.c..
\end{eqnarray}
$a^\dagger_{n}(b^\dagger_{n})$ and $a_{n}(b_{n})$ are the creation and annihilation operators at site-$n$ of the SSH chain. $t_1=t(1-\delta\cos \theta)$ represents the intracell hopping amplitudes, $t_2=t(1+\delta\cos \theta)$ is the intercell hopping amplitudes. $0<\delta<1$ is the modulation intensity, which describes the strength of dimerization. $\theta$ is the phase parameter and varies in one period, i.e., from $-\pi$ to $\pi$.
\par
In Eq.(1), the second term $U$ describes the onsite anti-$\cal PT$ symmetric imaginary potentials, which is achieved by introducing energy gain and loss, respectively. In this work, we would like to consider the application manner of the on-site imaginary potentials as ($i\gamma_1$, $-i\gamma_2$, $-i\gamma_1$, $i\gamma_2$). Obviously, the introduce of such on-site imaginary potentials causes our system to be the one-dimensional tetramerized lattice with four sublattices $A, B, C$ and $D$ [see Fig.1(b)]. According to our assumption, the Hamiltonian $U$ is expressed as
\begin{eqnarray}
U&=&\sum_{n}(i\gamma_1{a^\dag_{n} a_{n}}-i\gamma_2{b^\dag_{n} b_{4n-2}}\notag\\
&-&i\gamma_1{c^\dag_{n} c_{n}}+i\gamma_2{d^\dag_{n} d_{n}}),
\end{eqnarray}
where $\gamma_1$ and $\gamma_2$ are the strengths of two types of imaginary potentials. In the context, we consider them to be greater than zero.
\par
Following the Hamiltonian in Eqs.(1)-(2), we discuss the band structure and topological properties of this non-Hermitian SSH system. Under periodic boundary conditions, the Hamiltonian in the momentum space can be written by performing the Fourier transformation $\alpha_k=(1/\sqrt{N_c})\sum_k\alpha_n e^{ikn}$, that is,
\begin{eqnarray}
H=\sum_{k}{\psi^\dagger_{k}}H_{k}{\psi_{k}}.
\end{eqnarray}
Considering $\psi_{k}=[{a_k}, {b_k}, {c_k}, {d_k}]^T$, the expression of $H_k$ is given as
\begin{eqnarray}
H_{k}=
\left[\begin{array}{cccc}
i\gamma_1&t_1&0&{t_2}e^{-ik} \\
t_1&-i\gamma_2&t_2&0 \\
0&t_2&-i\gamma_1&t_1\\
{t_2}e^{ik}&0&t_1& i\gamma_2
\end{array}\right],
\end{eqnarray}
in which $k$ is the Bloch wavevector. This expression helps us to discuss in an analytical way the symmetry and phase transition conditions of our considered system.

\subsection{Symmetry}
The Hamiltonian symmetry determines the topological phases of symmetric protection in systems with topological structures. Thus, we first concentrate on the symmetry to present the topological properties of the non-Hermitian tetramerized SSH lattice.
\par
Due to our application manner of the imaginary potentials, $H_k$ has an opportunity to display the anti-$\cal PT$-symmetry, i.e., $({\cal{PT}})H_{k}({\cal{TP}})=-H_k$ with ${\cal{P}}=i\sigma_x \otimes \sigma_y$ and $\cal{T}=\kappa$. $\kappa$ corresponds to the complex-conjugation operation. In non-Hermitian physics, pseudo-Hermitianity is another important symmetry. By calculation, we find that the Hamiltonian of $H_{k}$ has pseudo-Hermiticity, i.e., $\eta H^{\dag}_k \eta^{-1}=H_k$, with the operator $\eta$ expressed as
\begin{eqnarray}
\eta=
\left[\begin{array}{cccc}
0&0&e^{-ik/2}&0 \\
0&0&0&e^{-ik/2} \\
e^{ik/2}&0&0&0\\
0&e^{ik/2}&0&0
\end{array}\right].
\end{eqnarray}
Surely, $\eta$ satisfies the conditions $\eta \eta^{\dagger}=\eta^{\dagger}\eta=1$ and $\eta^{\dagger}=\eta$. According to the previous works\cite{PRL}, this pseudo-Hermiticity operator can also be rewritten as ${\eta}=\sigma_{x}\otimes(\cos{k\over2})I_2+\sigma_{y}\otimes(\sin{k\over2})I_2$, where $\sigma_{x}$ and $\sigma_{y}$ are the Pauli matrices acting on the particle-hole representation. $I_2$ is the identity matrix. The pseudo-Hermitianity of Hamiltonian indicates the existence of anti-linear symmetry of the system. This shows the existence of a pure real part of the energy spectrum in this system.
\par
After the work pioneered by Kawabata $et$ $al.$\cite{PRX}, researchers tried to give definitions and conditions of various symmetries for non-Hermitian systems. For instance, particle-hole symmetry ($\mathrm{PHS}$) is defined as ${\cal{C}}_{-} H^T_k {\cal{C}}_{-}^{-1}=-H_{-k}$, time-reversal symmetry ($\mathrm{TRS}$) is defined as ${\cal{T}}_{+} H^*_k {\cal{T}}_{+}^{-1}=H_k$. The system also has $\mathrm{TRS}^{\dag}: {\cal{C}}_{+} H^T_k {\cal{C}}_{+}^{-1}=H_{-k}$ and $\mathrm{PHS}^{\dag}: {\cal{T}}_{-} H^*_k {\cal{T}}_{-}^{-1}=-H_{-k}$. From the expression of symmetry conditions, $\mathrm{TRS}^{\dagger}$ and $\mathrm{PHS}^{\dag}$ are defined as Hermitian conjugations of $\mathrm{TRS}$ and $\mathrm{PHS}$.
\par
For our system, the system does not satisfy $\mathrm{PHS}$ and $\mathrm{TRS}$ due to the presence of imaginary on-site potentials. However, we find that $H_{k}$ can satisfy $\mathrm{TRS}^{\dag}$ where the operator obeys ${\cal{C}}_{+}= I_2 \otimes I_2$ and ${\cal{C}_{+}\cal{C}_{+}^{*}}=1$. Similarly, $H_{k}$ has $\mathrm{PHS}^{\dag}$ symmetry and the operator ${\cal{T}}_{-}$ is expressed as
\begin{eqnarray}
{\cal{T}}_{-}=
\left[\begin{array}{cccc}
1&0&0&0 \\
0&-1&0&0 \\
0&0&1&0\\
0&0&0&-1
\end{array}\right]
\end{eqnarray}
and ${\cal{T}}_{-}{\cal{T}}_{-}^*=1$.
\par
In addition, it can be observed that $H_k$ satisfies pseudo-anti-Hermiticity (chiral symmetry). The chiral symmetry($\mathrm{CS}$) condition is $\Gamma H_k^\dag \Gamma^{-1}=-H_k$ where $\Gamma=\mathrm{diag}(1,-1,1,-1)$. As a combination of $\mathrm{TRS}^{\dag}$ and $\mathrm{PHS}^{\dagger}$, the chiral symmetry operator can be written as the anti-unitary operator $\Gamma=\cal{C}_{+}{\cal{T}}_{-}$. Meanwhile, the Hamiltonian of our system still satisfies $\mathrm{CS}^{\dagger}$ symmetry with ${\cal{C}} H_k {\cal{C}}^{-1}=-H_k$. The operator $\cal{C}$ satisfies
\begin{small}
\begin{eqnarray}
\cal{C}=
\left[\begin{array}{cccc}
0&0&e^{-ik/2}&0 \\
0&0&0&e^{-ik/2} \\
e^{ik/2}&0&0&0\\
0&e^{ik/2}&0&0
\end{array}\right]
\left[\begin{array}{cccc}
1&0&0&0 \\
0&-1&0&0 \\
0&0&1&0\\
0&0&0&-1
\end{array}\right].
\end{eqnarray}
\end{small}
$\mathrm{CS}^{\dagger}$ symmetry guarantees couples of eigenvalues $E(k)$. Therefore, by judging the symmetry, we can determine that our model belongs to the $\mathrm{BDI}^{\dag}$ category in the $38$-fold topological classifications of non-Hermitian systems\cite{PRXnn}. In Table I, the symmetries of the $\mathrm{BDI}^{\dag}$ class of non-Hermitian Hamiltonian is presented to further understand the symmetry of our system. Based on the previous works, the topological phase transition condition of the $\mathrm{BDI}^{\dag}$ class is determined by the band gap closing-reopening of the real part of energy bands\cite{RMP1}.
\par
\begin{table}[htb]
\caption{The $\mathrm{BDI}^{\dag}$ classes for non-Hermitian Hamiltonians.}
\begin{tabular}{p{1.7cm}p{1.3cm}p{0.85cm}p{0.85cm}p{0.85cm}p{0.85cm}p{0.5cm}}
\hline \hline
Symmetry&  &TRS&PHS&$\mathrm{TRS}^\dag$&$\mathrm{PHS}^\dag$&CS\\
Class&   &($\cal{T}_{+}$)&($\cal{C}_{-}$)&($\cal{C}_{+}$)&($\cal{T}_{-}$)&($\Gamma$) \\
\hline
Real ${\mathrm{AZ}}^{\dag}$&${\mathrm{BDI}}^{\dag}$&0&0&{+1}&{+1}&1 \\
\hline\hline
\end{tabular}
\end{table}
\begin{figure}
\begin{center}\scalebox{0.19}{\includegraphics{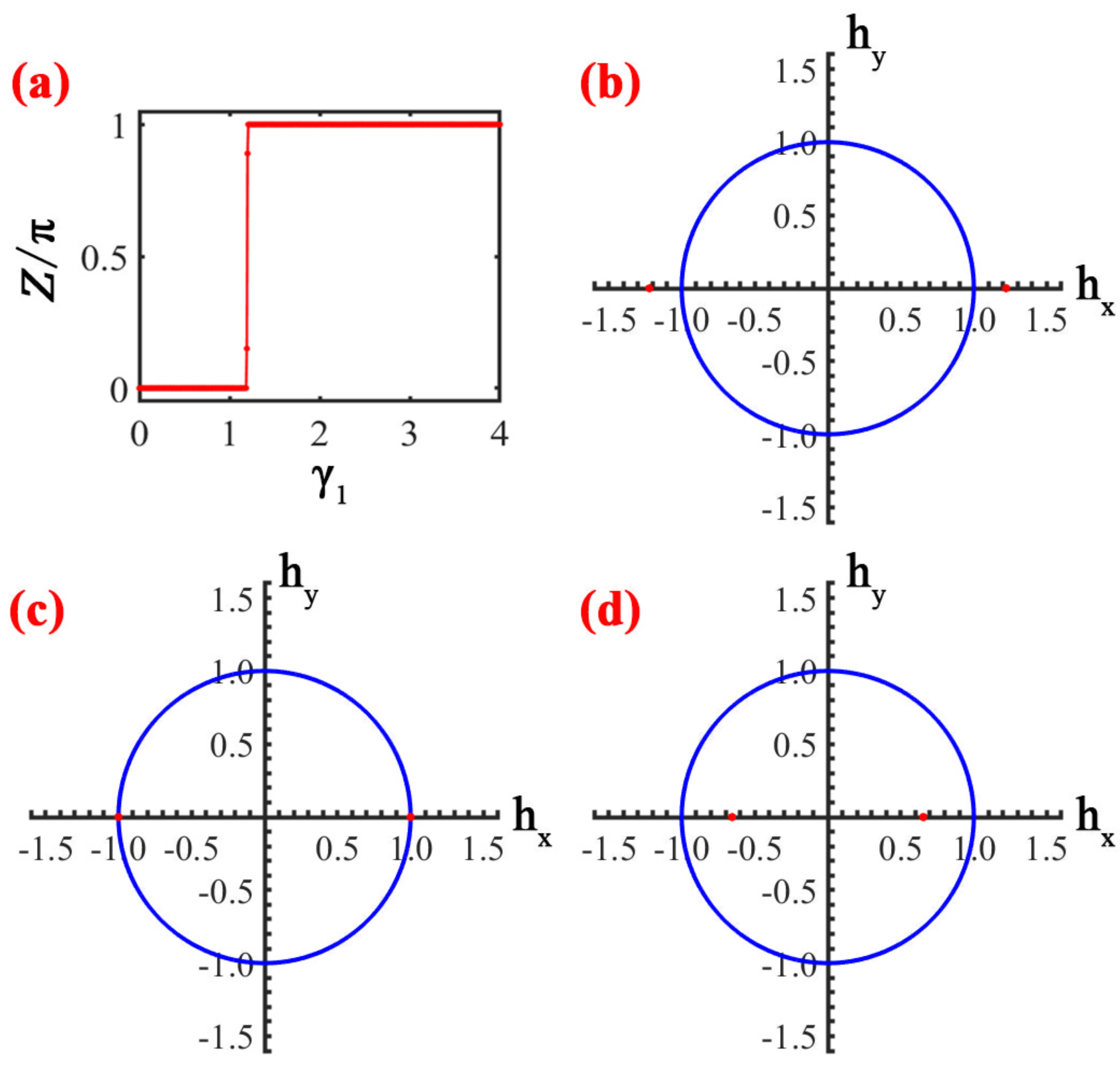}}
\caption{(a) Zak phase of $\gamma_1$. The parameters are $\gamma_2=1.0$, $\theta=\pi$ and $\delta=0.3$. (b)-(d) Geometrical picture of the winding number $W$ in the system around the topological transition. (b) $\gamma_1=0.6$, (c) $\gamma_1=1.2$ and (d) $\gamma_1=1.5$.}
\end{center}
\end{figure}
\subsection{Phase transition condition}

In this subsection, we would like to calculate the band structure of our SSH model, to further determine the topological phase transition conditions. By diagonalizing the Hamiltonian in Eq.(4), the relationship between the energy $E$, $\delta$, and $\theta$ can be obtained:
\begin{eqnarray}
E(k)=\pm{\frac{1}{\sqrt{2}}}\sqrt{X\pm  \sqrt{X^2-Y^2-16{t_1^2}{t_2^2}\sin^2{\frac{k}{2}}}},
\end{eqnarray}
where
\begin{eqnarray}
X&=&2(t_1^2+t_2^2)-\gamma_1^2-\gamma_2^2\notag \\
&=&4t^2(1+\delta^2\cos^2\theta)-\gamma_1^2-\gamma_2^2,\notag \\
Y&=&2[\gamma_1\gamma_2-(t_1^2-t_2^2)]=2\gamma_1\gamma_2+8t^2\delta\cos\theta.
\end{eqnarray}

When the imaginary potentials are equal to zero ($\gamma_1=\gamma_2=0$), our model is the Hermitian system, degeneralized into the standard SSH model. The eigenvalue expressions in the momentum space are simplified as
\begin{eqnarray}
E(k)=\pm \sqrt{t_1^2+t^2_2+2t_1t_2\cos k}.
\end{eqnarray}
In Fig.1(c)-(d), we plot the corresponding band structures under the periodic boundary condition of $\theta=0.5\pi$ and $\theta=0.8\pi$, respectively. When $t_1=t_2$, i.e., $\theta=0.5\pi$, the energy gap closes at $k=\pm\pi$, leading to the topological phase transition. According to the previous works, when $t_1\neq t_2$, the energy gap opens at $k=\pi$. Thus, as $\theta\in [-\pi, -0.5\pi]$ or $[0.5\pi, \pi]$, the system is located in the topologically-trivial phase. Otherwise, the system is topologically non-trivial.  

When the imaginary on-site potentials are not equal to zero ($\gamma_1\neq 0, \gamma_2\neq 0$), the model becomes a non-Hermitian system. Similarly, it is necessary to find where the energy gap closes and the phase transition occurs. We write out the expressions of the four bands as
\begin{small}
\begin{eqnarray}
E_1(k)&=&-{\frac{1}{\sqrt{2}}}\sqrt{X-\sqrt{X^2-Y^2-16{t_1^2}{t_2^2}\sin^2{\frac{k}{2}}}}, \notag\\
E_2(k)&=&{\frac{1}{\sqrt{2}}}\sqrt{X-\sqrt{X^2-Y^2-16{t_1^2}{t_2^2}\sin^2{\frac{k}{2}}}}, \notag\\
E_3(k)&=&-{\frac{1}{\sqrt{2}}}\sqrt{X+\sqrt{X^2-Y^2-16{t_1^2}{t_2^2}\sin^2{\frac{k}{2}}}}, \notag\\
E_4(k)&=&{\frac{1}{\sqrt{2}}}\sqrt{X+\sqrt{X^2-Y^2-16{t_1^2}{t_2^2}\sin^2{\frac{k}{2}}}}.
\end{eqnarray}
\end{small}
If the bands meet each other, $\gamma_1$, $\gamma_2$, $t_1$ and $t_2$ should satisfy $X^2-Y^2-16t_1^2t_2^2\sin^2{\frac{k}{2}}=0$. For example, when $k=0$, we can get $X=\pm Y$. There are two situations for $X$ and $Y$. When $X=Y$, we obtain the relationship between $\gamma_1$, $\gamma_2$ and $t_1$ manifested as $\gamma_2=2t_1-\gamma_1$ and $\gamma_2=-2t_1-\gamma_1$. Since we consider $\gamma_1>0$ and $\gamma_2>0$, the case of less than zero is ignored. Under this condition, the imaginary part of the energy band $\mathrm{Im}(E)$ degenerates to be zero at $k=0$, while the real part is not equal to zero.
On the other hand, for $X=-Y$, the relationship of $\gamma_2=-2t_2+\gamma_1$ and $\gamma_2=2t_2+\gamma_1$ is obtained. At this point, the real part of the energy band $\mathrm{Re}(E)=0$ at $k=0$, while $\mathrm{Im}(E)\neq 0$.
\par
Next if $E_1$ and $E_2$ meet, the condition should be expressed as $X-\sqrt{X^2-Y^2-16{t_1^2}{t_2^2}\sin^2{\frac{k}{2}}}=0$. For $k=0$, it simplifies to $\gamma_1\gamma_2=t_1^2-t_2^2=-4t^2\delta\cos{\theta}$. Both the real and imaginary parts of $E(k)$ are equal to zero. According to the conclusion of symmetry of $\mathrm{BDI}^{\dag}$ class, the gap-closing conditions correspond to the occurrence of phase transition. Therefore, in order to ascertain the phase transition conditions, $\theta$ must obey the relationship of $\cos{\theta}<0$, which corresponds to the topologically-trivial region of the Hermitian SSH model.
\par
We can work out the eigenvectors of $H_k$ expressed as
\begin{small}
\begin{widetext}
\begin{eqnarray}
\Psi_1&=&\left(\frac{-A_2+B+C_1(C_1+iD_1)}{t_2E_1D_1-2it_2\gamma_1F_1},\frac{-4i(C_1t_{1}^2-C_2t_{2}^2)-(B-C_1C_2)(D_1+2i\gamma_2)}{2t_1t_2[E_1D_1-2i\gamma_1F_1]},\frac{-A_1+B+C_2(C_2-iD_1)}{t_1E_1D_1-2it_1\gamma_1F_1},1\right),\notag\\
\Psi_2&=&\left(\frac{A_2-B-C_1(C_1-iD_1)}{t_2E_1D_1+2it_2\gamma_1F_1},\frac{4i(C_1t_{1}^2-C_2t_{2}^2)+(B-C_1C_2)(-D_1+2i\gamma_2)}{2t_1t_2[E_1D_1+2i\gamma_1F_1]},\frac{A_1-B-C_2(C_2+iD_1)}{t_1E_1D_1+2it_1\gamma_1F_1},1\right),\notag\\
\Psi_3&=&\left(\frac{-A_2-B+C_1(C_1+iD_2)}{t_2E_1D_2-2it_2\gamma_1F_1},\frac{-4i(C_1t_{1}^2-C_2t_{2}^2)+(B+C_1C_2)(D_2+2i\gamma_2)}{2t_1t_2[E_1D_2-2i\gamma_1F_1]},\frac{-A_1-B+C_2(C_2-iD_2)}{t_1E_1D_2-2it_1\gamma_1F_1},1\right),\notag\\
\Psi_4&=&\left(\frac{A_2+B-C_1(C_1-iD_2)}{t_2E_1D_2+2it_2\gamma_1F_1},\frac{4i(C_1t_{1}^2-C_2t_{2}^2)-(B+C_1C_2)(-D_2+2i\gamma_2)}{2t_1t_2[E_1D_2+2i\gamma_1F_1]},\frac{A_1+B-C_2(C_2+iD_2)}{t_1E_2D_1+2it_1\gamma_1F_1},1\right). \end{eqnarray}
\end{widetext}
\end{small}
The relevant parameters are defined as $A_1=2t_{1}^2(1+e^{ik})$, $A_2=2t_{2}^2(1+e^{-ik})$, $B=\sqrt{X^2-Y^2-16{t_1^2}{t_2^2}\sin^2{\frac{k}{2}}}$, $C_1(C_2)=\gamma_1-(+)\gamma_2$, $D_1(D_2)=\sqrt{2(X-(+)B)}$, $E_1=1+e^{ik}$ and $F_1=e^{ik}-1$. Based on these eigenvectors, the topological properties of this system can be well distinguished by the Zak phase. Then we introduce the Zak phase\cite{z3,z4,z5} in the momentum space defined as ${\cal Z}=i\int^{4\pi}_{0}{\langle\varphi_n|\partial_k{\psi_n}\rangle dk}$. $|{\psi_n}\rangle$ is the right eigenstate and $|{\varphi_n}\rangle$ is left eigenstate, $H_k{\psi_n}=E_n{\psi_n}$ and $H_k^\dag{\varphi_n}=E_n^*{\varphi_n}$. By solving the Zak phase $\mathcal{Z}$, we can determine the topologically nontrivial and trivial region of the system. Fig.2(a) shows the relationship between $\mathcal{Z}$ and $\gamma_1$, where $\theta=\pi$, $\delta=0.3$ and $\gamma_2=1.0$. When $\gamma_1<1.2$, $\mathcal{Z}=0$, whereas if $\gamma_1>1.2$, $\mathcal{Z}=\pi$. Therefore, under the condition of $t_1>t_2$, we get integer values of $\mathcal{Z}=\pi$ for $\gamma_1\gamma_2>-4t^2\delta\cos{\theta}$ and $\mathcal{Z}=0$ for $\gamma_1\gamma_2<-4t^2\delta\cos{\theta}$.
\par
The topology of the non-Hermitian tetramerized SSH model also relates to the geometry of the Bloch Hamiltonian winding around degeneracy points in a two-dimensional parameter space. Next, we show how the non-Hermiticity creates the nontrivial topology. For the Bloch Hamiltonian Eq.(4), we replace $e^{ik}$ by $h_x+ih_y$ to create a two-dimensional parameter space $(h_x,h_y)$,
\begin{small}
\begin{widetext}
\begin{eqnarray}
H_{h_x,h_y}&=&
\left[\begin{array}{cccc}
i\gamma_1&t_1&0&{t_2}(h_x-ih_y)^2 \\
t_1&-i\gamma_2&t_2(h_x^2+h_y^2)&0 \\
0&t_2(h_x^2+h_y^2)&-i\gamma_1&t_1\\
{t_2}(h_x+ih_y)^2&0&t_1& i\gamma_2
\end{array}\right] \notag\\
&=&t_1(I_2\otimes\sigma_x)+t_2(h_x\sigma_x+h_y\sigma_y)\otimes(h_x\sigma_x+h_y\sigma_y)+i(\frac{\gamma_1+\gamma_2}{2})(\sigma_z\otimes\sigma_z)+i(\frac{\gamma_1-\gamma_2}{2})(\sigma_z\otimes I_2).
\end{eqnarray}
\end{widetext}
\end{small}
We set a unit circle $h_x^2+h_y^2=1$ in the two-dimensional parameter space ($h_x,h_y$). Based on the previous content, we can infer that the gain and loss imaginary on-site potentials adjust the degenerate zero-energy state. Since the Dirac point appears at $k=0$, we can omit the imaginary $\sigma_y$ term, i.e., $h_y=0$. By setting ${t_2}'=t_2(h_x^2+h_y^2)$ and diagonalizing the Hamiltonian, we get
\begin{eqnarray}
E(k)=\pm{\frac{1}{\sqrt{2}}}\sqrt{X'\pm \sqrt{Y'}},
\end{eqnarray}
where $X'=2({t_2}'+t_1^2)-\gamma_1^2-\gamma_2^2$, $Y'=[-{4{t_2}'}^2+(\gamma_1-\gamma_2)^2][(\gamma_1+\gamma_2)^2-4t_1^2]$. The resultant condition for degeneracy is given by $2({t_2}'+t_1^2)-\gamma_1^2-\gamma_2^2+\sqrt{[-{4{t_2}'}^2+(\gamma_1-\gamma_2)^2][(\gamma_1+\gamma_2)^2-4t_1^2]}=0$. After derivation, we obtain the formula:
\begin{eqnarray}
h_x^2=\sqrt{\frac{t_1^2-\gamma_1\gamma_2}{t_2^2}},
\end{eqnarray}
Thus, the degenerate zero points can be displaced to be
\begin{eqnarray}
(h_x,h_y)=\left(\pm\left(\frac{t_1^2-\gamma_1\gamma_2}{t_2^2}\right)^{\frac{1}{4}},0\right).
\end{eqnarray}
Fig.2(b)-(d) display the structure extended to the two-dimensional parameter space ($h_x,h_y$) for the non-Hermitian tetramerized SSH model at fixed parameters $\theta=\pi$, $\gamma_2=1.0$ and $\delta=0.3$ for the different non-Hermiticity $\gamma_1$. It is shown that when $\gamma_1=0.6$, the zero points (red point) move to ($\pm1.22,0$). Those are out of the winding, so the system is in the topologically-trivial region where $W=0$, as shown in Fig.2(b). In Fig.2(c), when $\gamma_1$ increases to $\gamma_1=1.0$, the zero points moves to $(h_x,h_y)=(\pm1,0)$ and locates on the unit circle. At this time, the system is located at the critical point of topological phase transition. For $\gamma_1=1.6$ in Fig.2(d), the zero points are enclosed in the unit circle, the topology of the system changes, and it enters the topologically nontrivial region where the winding number is equal to $1.0$. The results are consistent with the previous conclusion. Therefore, for $\theta\in [-\pi, -0.5\pi]$ or $[0.5\pi, \pi]$, when the $\gamma_1\gamma_2<t_1^2-t_2^2$, the system is topologically trivial. In the case of $\gamma_1\gamma_2>t_1^2-t_2^2$, the system enters the topologically nontrivial region. For $\theta\in [-0.5\pi, 0.5\pi]$, the system is always located in the topologically nontrivial region.

\section{Numerical results and discussions}
Following the theoretical deduction in Sec.II, we proceed to investigate the band structure of the non-Herimitian tetramerized SSH lattice, aiming at the clarification of the topological properties of it.

\begin{figure}
\begin{center}\scalebox{0.12}{\includegraphics{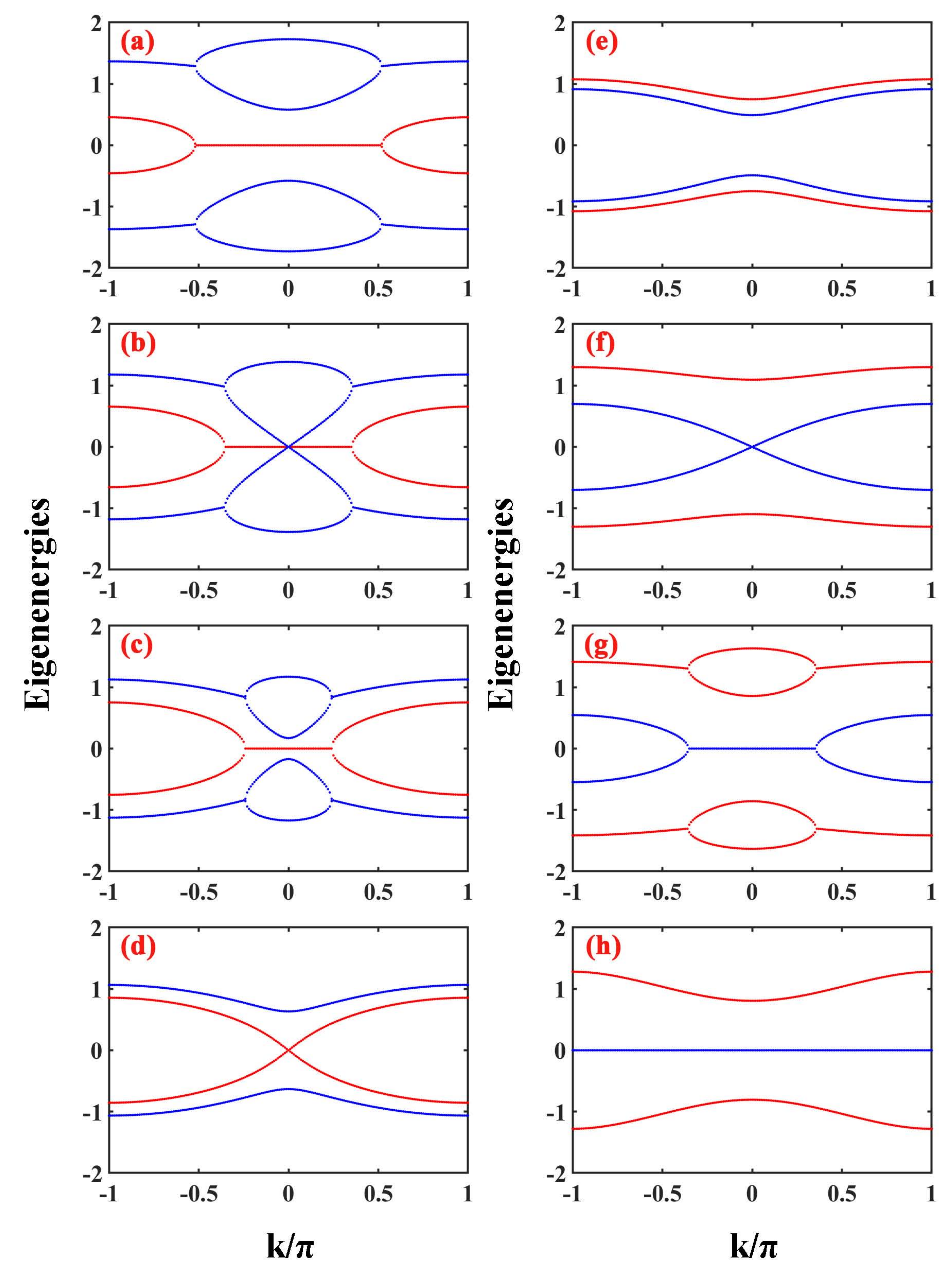}}
\caption{Band structures for different values of imaginary potential $\gamma_1$. The system parameters are taken to be $\delta=0.3$, $\theta=\pi$, and $\gamma_2=1.0$. The value of $\gamma_1$ is respectively set to be (a) $\gamma_1=0.5$, (b) $\gamma_1=1.2$, (c) $\gamma_1=1.4$, (d) $\gamma_1=1.6$, (e) $\gamma_1=2.0$, (f) $\gamma_1=2.4$, (g) $\gamma_1=2.6$, and (h) $\gamma_1=3.0$. The blue lines indicate the real part of energy, and the red lines correspond to the imaginary part.}
\end{center}
\end{figure}

\par
To begin with, we would like to plot the spectra of the eigenenergies of the SSH lattice in the momentum space, by considering different imaginary potentials. The numerical results are shown in Fig.3, in which the structural parameters are taken to be $t=1.0$, $\delta=0.3$, $\theta=\pi$, and $\gamma_2=1.0$. It can be found that with the increase of $\gamma_1$, this system undergoes different phase transition processes. This means the nontrivial role of the imaginary potentials in modifying the topological properties of the SSH chain. Firstly, in Fig.3(a)-(c) it shows that when $\gamma_1$ increases to 1.2, the second and third bands are allowed to encounter at the point of $k=0$, leading to the formation of the Dirac cone of the band structure [see Fig.3(b)]. However, the following increment of $\gamma_1$ destroys the Dirac cone and causes the reopening of the band gap, as shown in Fig.3(c). One can know that in such a process, the topological phase transition occurs. Next if $\gamma_1$ increases to 1.6, the real part of energy will degenerate into two branches, whereas the imaginary part of energy displays the Dirac cone around the point of $k=0$. This Dirac cone can be destroyed by the increase of $\gamma_1$ as well. In the following, the real and imaginary parts of energy display the similar variation manner [see Fig.3(d)-(e)]. This process describes the phase transition of the anti-$\cal PT$ symmetry. However, in Fig.3(f)-(h) it can be found that when $\gamma_1\ge2.4$, the real part of energy has another opportunity to meet at the point of $k=0$. The further increase of $\gamma_1$ will extend the region of zero real part of energy. As a result, when $\gamma_1=3.0$, the real part of energy keeps equal to zero in the whole region. In view of these results, we understand that in the SSH chain, our considered non-Hermitian imaginary potentials can induce the abundant phase transitions.
\begin{figure}
\begin{center}\scalebox{0.12}{\includegraphics{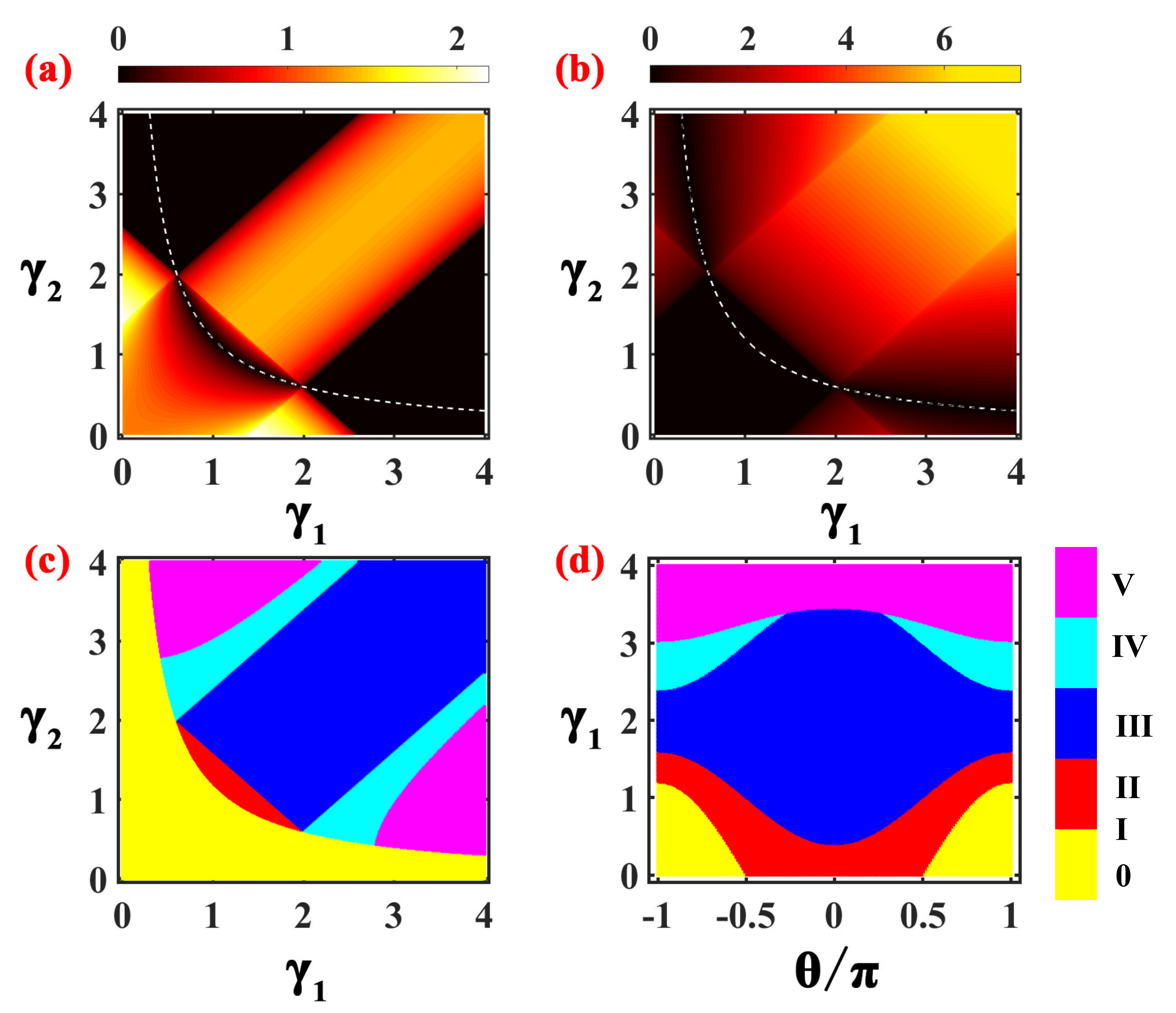}}
\caption{(a)-(b) Schematic representing the phase diagrams for the band gap with the change of $\gamma_1$ and $\gamma_2$. (a) Real part of the eigenvalues $E(k)$. (b) Imaginary part of the eigenvalues $E(k)$. The legend represents the width of the gap between the central energy bands. (c)-(d) Schematic representing the phase diagram for the band structure of our system with (c) the change of $\gamma_1$ and $\gamma_2$, when $\theta=\pi$ and $\delta=0.3$, (b) the change of $\theta$ and $\gamma_1$, when $\gamma_2=1.0$ and $\delta=0.3$.}
\end{center}
\end{figure}
\par
In order to present the phase transition properties of this tetramerized non-Hermitian SSH lattice, we plot the phase diagram of them. For comparison, Fig.4(a)-(b) show the widths of the band gaps for $\mathrm{Re} E(k)$ and $\mathrm{Im} E(k)$ depending on the gain and loss. One can clearly find that the gain and loss efficiently modulate the band gaps of the real and imaginary parts of the energy band. As shown in Fig.4(a), the finite band gap of $\mathrm{Re} E(k)$ mainly exists in two regions, i.e., the region of $\gamma_2+\gamma_1\le2.5$ and $\gamma_1-1.5\le\gamma_2\le\gamma_1+1.5$. In the other regions, the band gap of $\mathrm{Re} E(k)$ is closed. When observing the result in Fig.4(b), we see that the band gap of $\mathrm{Im} E(k)$ displays the similar change to the real part $\mathrm{Re} E(k)$. The difference is that after exceeding the critical lines, the band gap of $\mathrm{Im} E(k)$ decreases gradually to zero.
Following this result, we present the diagram of the phase transition of our system in Fig.4(c)-(d). In order to compare the relation between the topological phase transition and the band gap changes, we plot the phase transition diagram with the change of $\gamma_1$ and $\gamma_2$. As a result, the phase diagram is divided into five regions. The yellow region corresponds to the topologically trivial region, and the other part describes the topologically nontrivial region. In the latter, we use four colors to denote the different non-Hermitian regions. The red color means the zero $\mathrm{Im} E(k)$, whereas the dark blue region describes the complex $E(k)$. These results are well consistent with those in Fig.3(c)-(d). For the light blue region, it can be related to Fig.3(g), whereas the pink  region can be regarded as the description of Fig.3(h). Next, Fig.4(d) shows the co-influence of $\gamma_1$ and $\theta$ on the phase transition. It also tells us that the anti-$\cal PT$ symmetry transition only occurs in the topologically nontrivial region.

\begin{figure}[htb]
\begin{center}\scalebox{0.12}{\includegraphics{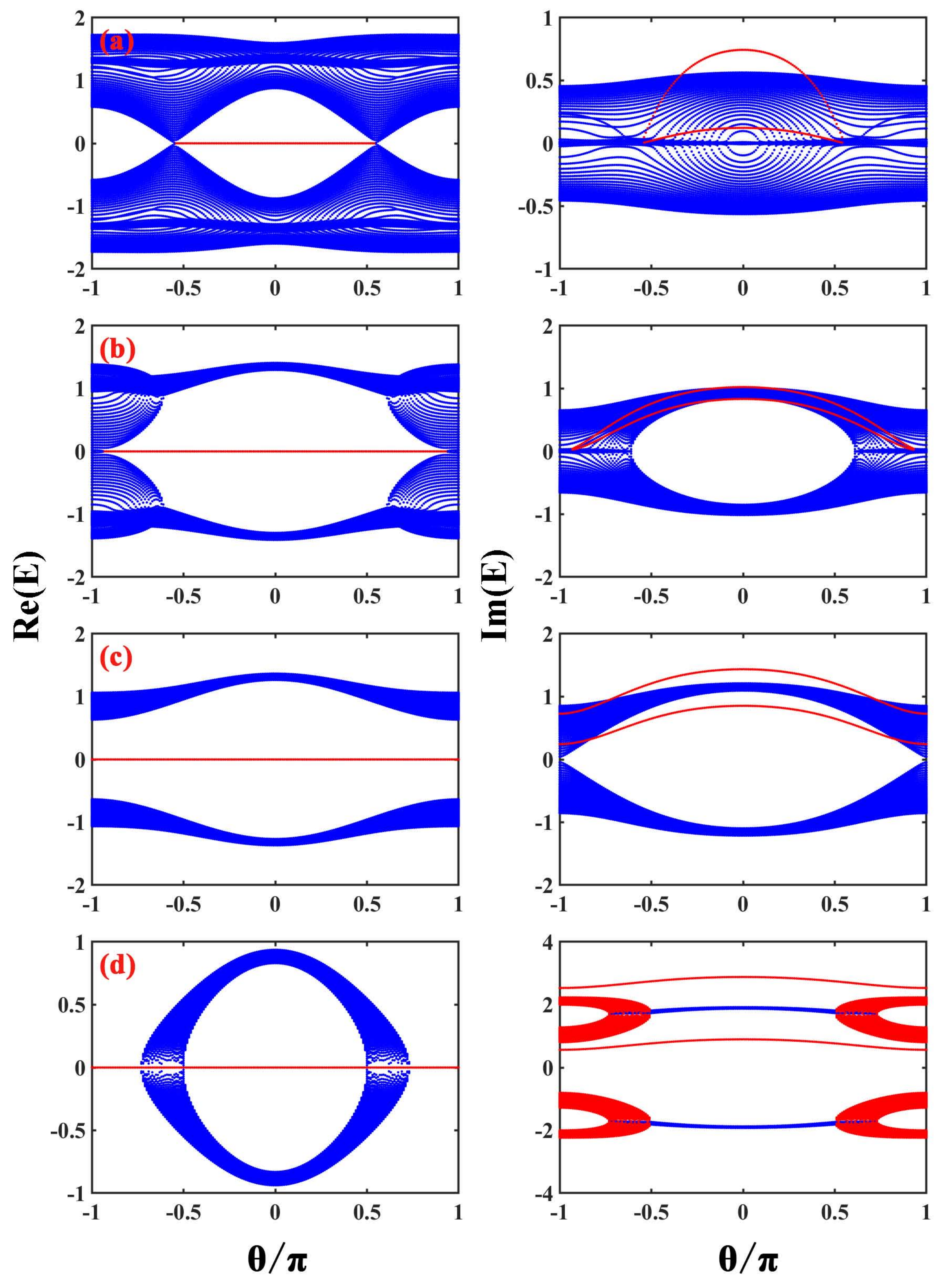}}
\caption{Real and imaginary parts of energy for different $\theta$. The parameters are (a) $\gamma_1=0.2$, (b) $\gamma_1=1.2$, (c) $\gamma_1=1.6$, and (d) $\gamma_1=3.0$. Left panel shows the real part of energy, and the right corresponds to the imaginary part of energy. Other parameters are set to be $\gamma_2=1.0$ and $\delta=0.3$.}
\end{center}
\end{figure}
\begin{figure}[htb]
\begin{center}\scalebox{0.10}{\includegraphics{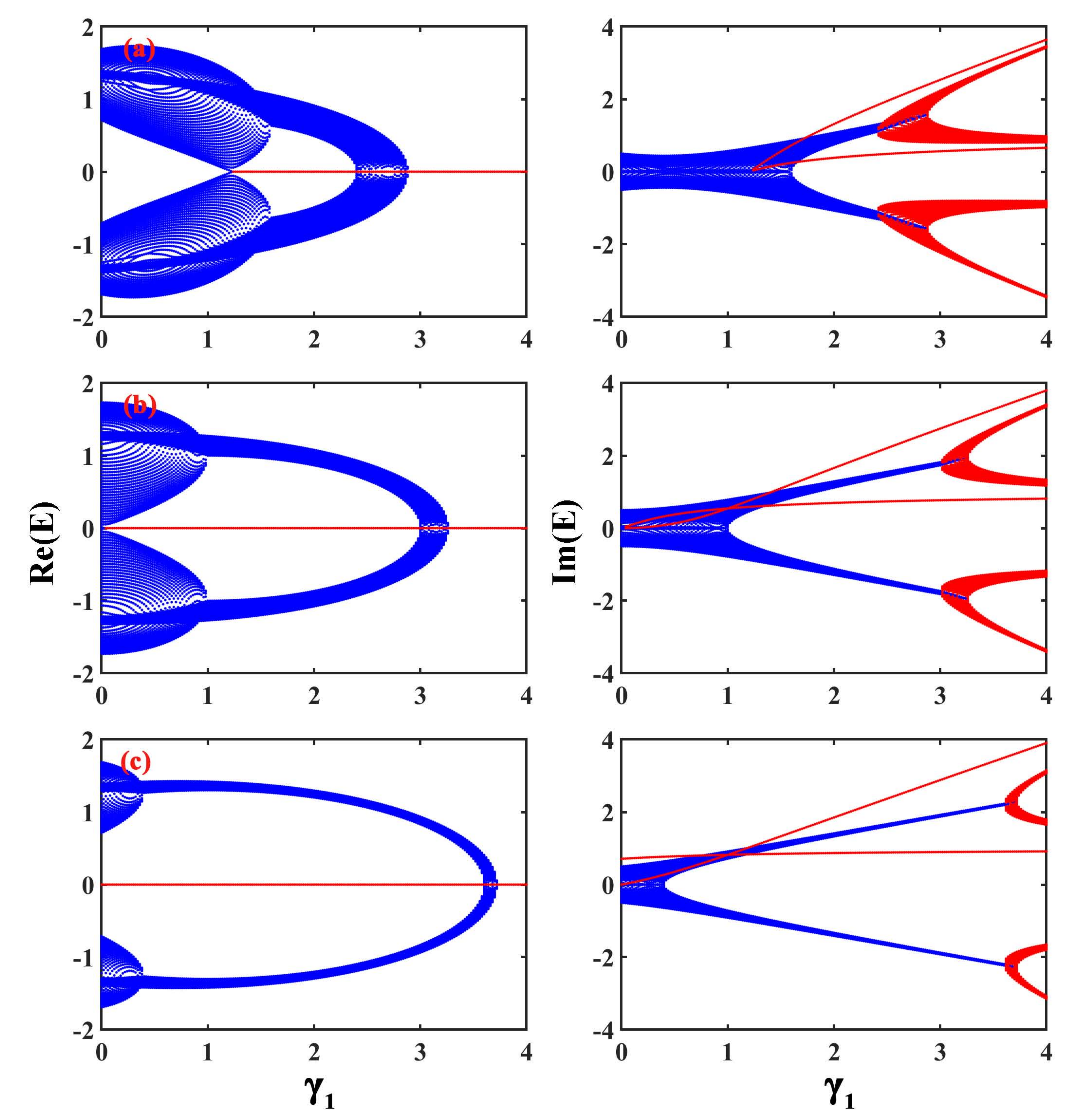}}
\caption{Results of real and imaginary energy spectra in different topological phase regions, with the increase of $\gamma_1$. Parameter $\theta$ is taken to be: (a) $\theta=\pi$, (b) $\theta=0.5\pi$, and (c) $\theta=0$. Other parameters are given as $\gamma_2=1.0$ and $\delta=0.3$.}
\end{center}
\end{figure}
\par
Next, we turn to the case of a finite-length SSH chain to present the appearance of the topological states. Fig.5 shows the real and imaginary parts of energies under the condition of $N_c=50$. The other structural parameters are set to $\gamma_2=1.0$ and $\delta=0.3$. In Fig.5(a) where $\gamma_1=0.2$, the real part of energy displays the two-band structure, and these two bands encounter at the points of $\theta=\pm0.5\pi$. Between these two phase transition points, zero-energy zero modes appear. It is notable that the imaginary parts of the bulk-state energies are symmetric around their zero value, but for the topological states they are both positive. This phenomenon is completely different from the $\cal PT$-symmetric case. When $\gamma_1$ increases to 1.2, the topologically-nontrivial region is greatly widened, accompanied by narrowness of the bulk-state bands. As the value of $\theta$ approaches $\pi$, the bulk-state bands are widened and cover the zero-energy modes. For $\gamma_1=1.6$, the bulk-state bands are further narrowed in the whole region, and then no energy crossing occurs. Note that in this case, the imaginary parts of the topological-state energies keep located in the positive-value region with their in-phase oscillation. Next, in Fig.5(d), for the extremum case of $\gamma_1=3.0$, the zero-energy modes exist in the whole region, but the bulk-states appear in the region of $-0.75\pi<\theta<0.75\pi$. Around the points of $\theta=\pm0.5\pi$, the two bands merge into one, which means the nontrivial transition of the band structure of our system. Therefore, the band structure and the topological states can be adjusted efficiently in this SSH chain.

\begin{figure*}[htb]
\begin{center}\scalebox{0.11}{\includegraphics{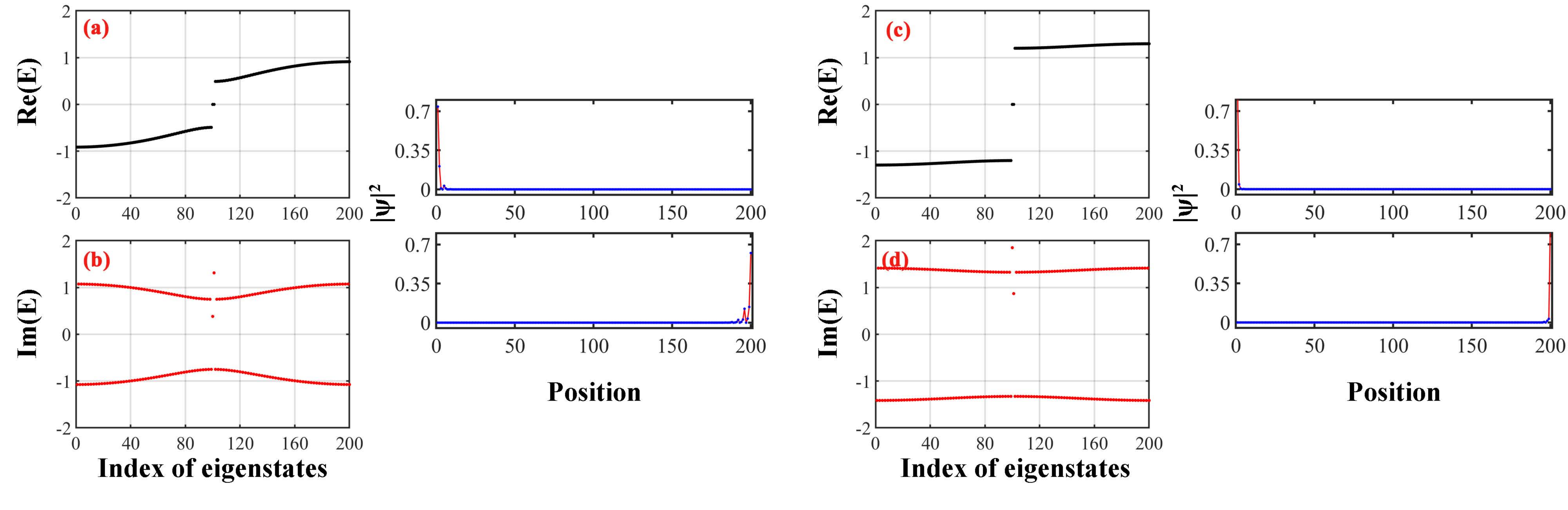}}
\caption{Energy and LDOS spectra with open boundary conditions with $\gamma_1=2.0$, $\delta=0.3$ and $\gamma_2=1.0$. The left is the real and imaginary parts of energy with (a)-(b) $\theta=\pi$, and (c)-(d) $\theta=0$.}
\end{center}
\end{figure*}
\par
Fig.6 shows the energy spectra modulated by the increase of $\gamma_1$, where $\gamma_2=1.0$ and $\delta=0.3$. In Fig.6(a)-(c), we set $\theta=\pi$, $0.5\pi$, and $0$ to present the influence of $\gamma_1$ on the topological phase transition. Firstly, Fig.6(a) displays that with the increase of $\gamma_1$ to 1.2, the bulk-state bands meet at the energy zero point, thus inducing the appearance of the topological states. This means that $\gamma_1$ plays an important role in changing the topological properties of the SSH lattice. Further increasing in $\gamma_1$ causes a severe narrowness of the bulk bands and then closes the band gap. However, the topological states are robust in this process. Next, when $\theta=0.5\pi$, the increase of $\gamma_1$ directly drives the appearance of topological states. The following change of the bulk states is similar to $\theta=\pi$. When comparing the imaginary parts of energy in Fig.6(a)-(b), we see that in the latter case, the imaginary energy of the topological states are degenerated at the critical point of $\gamma_1=1.0$. As for $\theta=0$, the main effect of the imaginary potential is to change the bulk bands and the imaginary part of energy. The real part of topological-state energy always exist at the energy zero point. The imaginary energies of the topological states are also degenerated at the critical point of $\gamma_1=1.0$. In view of these results, the degeneracy of the imaginary topological-state energy is mainly determined by $\gamma_2$.

\par
To determine the characteristics of edge states in this tetramerized non-Hermitian SSH system, we present the numerical analysis of the energy spectra and local density of states (LDOS) in the real space under open boundary conditions. The corresponding results are shown in Fig.7, where respective parameters are $\gamma_1=2.0$, $\delta=0.3$ and $\gamma_2=1.0$. On the one hand, Fig.7(a) describes the case where $\theta=\pi$, whereas Fig.7(b) corresponds to the case of $\theta=0$. The size of the lattice is set to $N_c=100$. In Fig.7(a), the real and imaginary parts of the energy spectrum display the two-band structure with a narrow band gap in between. For the real part, the topological states are clearly present in the gap with their zero-energy levels. The imaginary part of the energy spectrum displays its alternative results. Specifically, the bulk-state part is symmetric about the energy zero point, but the topological-state part is manifested as two positive imaginary energies. From the LDOS profiles, each topological state is localized at one end of the lattice, respectively. On the other hand, for the case of $\theta=0$, the similar phenomenon is shown in Fig.7(b). Compared to $\theta=\pi$, the band widths are greatly decreased, whereas the band gap is greatly enlarged. Meanwhile, the imaginary part of the energy spectrum exhibits a much larger magnitude. Also in this case, the imaginary part of the topological states distributes in the positive-value region. From the LDOS results, one can see that the LDOS spectra of the topological states are more localized compared to $\theta=\pi$. Until now, we have better comprehended the topological properties in the tetramerized non-Hermitian SSH system.
\par

\section{summary}
In summary, we have discussed the topological properties in the non-Hermitian SSH lattice by means of the periodical introduction of the onsite imaginary potentials ($i\gamma_1$, $-i\gamma_2$, $-i\gamma_1$, $i\gamma_2$). We found that by inducing the change of the SSH lattice to the tetramerized non-Hermitian system, this system displays new and interesting symmetries. As a result, such imaginary potentials lead to the nontrivial transition in the topological properties of the tetramerized SSH chain. On the one hand, the topologically-nontrivial region is extended, followed by the spontaneous breaking of the anti-$\cal PT$ symmetry. In this process, new edge state occurs, but its locality is different from that in the Hermitian SSH lattice. Moreover, if the imaginary potentials are strong enough, the bulk states of this system can be thoroughly imaginary. These imaginary potentials play special roles in modulating the topological properties of the non-Hermitian SSH lattice. We believe that our findings contribute to the understanding of the topological phase transition behaviors in the anti-$\cal PT$-symmetric non-Hermitian system. Moreover, due to the recent experimental advantages in non-Hermitian physics field, we anticipate that the results in this work can be realized in optical-lattice systems by introducing gain and loss.

\section*{Acknowledgments}
\par

This work was financially supported by the LiaoNing Revitalization Talents Program (Grant No. XLYC1907033), the National Natural Science Foundation of China (Grant No. 11905027), the Funds for the Central Universities (Grants No. N2002005).

\clearpage

\bigskip

\end{document}